\documentclass{aa}
\usepackage[varg]{txfonts}
\usepackage{graphicx}
\usepackage{natbib,twoopt}
\usepackage[breaklinks=true]{hyperref} 
\usepackage{lscape}
\bibpunct{(}{)}{;}{a}{}{,} 
\makeatletter
 \newcommandtwoopt{\citeads}[3][][]{\href{https://ui.adsabs.harvard.edu/abs/#3/abstract}%
 {\def\hyper@linkstart##1##2{}%
 \let\hyper@linkend\@empty\citealp[#1][#2]{#3}}}
 \newcommandtwoopt{\citepads}[3][][]{\href{https://ui.adsabs.harvard.edu/abs/#3/abstract}%
 {\def\hyper@linkstart##1##2{}%
 \let\hyper@linkend\@empty\citep[#1][#2]{#3}}}
 \newcommandtwoopt{\citetads}[3][][]{\href{https://ui.adsabs.harvard.edu/abs/#3/abstract}%
 {\def\hyper@linkstart##1##2{}%
 \let\hyper@linkend\@empty\citet[#1][#2]{#3}}}
 \newcommandtwoopt{\citeyearads}[3][][]%
 {\href{https://ui.adsabs.harvard.edu/abs/#3/abstract}
 {\def\hyper@linkstart##1##2{}%
 \let\hyper@linkend\@empty\citeyear[#1][#2]{#3}}}
\makeatother

\begin{document}

   \title{When the horseshoe fits: Characterizing 2023~FY$_{3}$ with the 10.4~m Gran Telescopio Canarias and the Two-meter Twin Telescope\thanks{Based 
          on observations made with the Gran Telescopio Canarias (GTC) telescope, in the Spanish Observatorio del Roque de los Muchachos of the 
          Instituto de Astrof\'{\i}sica de Canarias (IAC, program ID GTC31-23A) and the Two-meter Twin Telescope (TTT), in the Spanish Observatorio 
          del Teide of the IAC (commissioning phase).}}
   \author{R.~de~la~Fuente Marcos\inst{1}
            \and
           C.~de~la~Fuente Marcos\inst{2}
            \and
           J.~de~Le\'on\inst{3,4}
            \and 
           M.~R. Alarcon\inst{3,4}
            \and 
           J. Licandro\inst{3,4}
            \and
           M. Serra-Ricart\inst{3,4,5}
            \and
           D. Garc\'{\i}a-\'Alvarez\inst{6,3}
            \and
           A. Cabrera-Lavers\inst{6,3,4}
          }
   \authorrunning{R. de la Fuente Marcos et al.}
   \titlerunning{Physical characterization of 2023~FY$_{3}$ with GTC and TTT} 
   \offprints{R. de la Fuente Marcos, \email{rauldelafuentemarcos@ucm.es}}
   \institute{AEGORA Research Group,
              Facultad de Ciencias Matem\'aticas,
              Universidad Complutense de Madrid,
              Ciudad Universitaria, E-28040 Madrid, Spain
              \and
              Universidad Complutense de Madrid,
              Ciudad Universitaria, E-28040 Madrid, Spain
              \and
              Instituto de Astrof\'{\i}sica de Canarias (IAC),
              C/ V\'{\i}a L\'actea s/n, E-38205 La Laguna, Tenerife, Spain
              \and
              Departamento de Astrof\'{\i}sica, Universidad de La Laguna,
              Avda. Astrof\'{\i}sico Francisco S\'anchez, E-38206 La Laguna, Tenerife, Spain
              \and
              Light Bridges S.L., 
              Avda. Alcalde Ram\'{\i}rez Bethencourt, 17, E-35004, Las Palmas de Gran Canaria, Canarias, Spain
              \and
              GRANTECAN,
              Cuesta de San Jos\'e s/n, E-38712 Bre\~na Baja, La Palma, Spain
             }
   \date{Received 7 August 2023 / Accepted 12 October 2023}
   \abstract
      {The Arjuna asteroid belt is loosely defined as a diverse group of small 
       asteroids that follow dynamically cold, Earth-like orbits. Most of them 
       are not actively engaged in resonant, co-orbital behavior with Earth. 
       Some of them experience temporary but recurrent horseshoe episodes. 
       Objects in horseshoe paths tend to approach Earth at a low velocity, leading 
       to captures as Earth's temporary satellites or mini-moons. Four such 
       objects have already been identified: 1991~VG, 2006~RH$_{120}$, 
       2020~CD$_{3}$, and 2022~NX$_{1}$. Here, we focus on 2023~FY$_{3}$, a 
       recent finding, the trajectory of which might have a co-orbital status and perhaps 
       lead to temporary captures.  
      }
      {We want to determine the physical properties of 2023~FY$_{3}$ and explore 
       its dynamical evolution.
       }
      {We carried out an observational study of 2023~FY$_{3}$ using the OSIRIS 
       camera spectrograph at the 10.4~m Gran Telescopio Canarias, to derive its 
       spectral class, and time-series photometry obtained with QHY411M cameras 
       and two units of the Two-meter Twin Telescope to investigate its 
       rotational state. $N$-body simulations were also performed to examine its 
       possible resonant behavior.
       }
      {The visible reflectance spectrum of 2023~FY$_{3}$ is consistent with that 
       of an S-type asteroid; its light curve gives a rotation period of 
       9.3$\pm$0.6~min, with an amplitude of 0.48$\pm$0.13~mag. We confirm that 
       2023~FY$_{3}$ roams the edge of Earth's co-orbital space. 
       }
      {Arjuna 2023~FY$_{3}$, an S-type asteroid and fast rotator, currently 
       exhibits horseshoe-like resonant behavior and in the past experienced mini-moon 
       engagements of the temporarily captured flyby type that may 
       repeat in the future. The spectral type result further confirms that 
       mini-moons are a diverse population in terms of surface composition.
       }


   \maketitle

   \section{Introduction\label{Intro}}
      \citet{1993Natur.363..704R} pointed out that the path followed by the Earth--Moon system is strewn with debris in the form of small asteroids. 
      In fact, the heliocentric orbits of these objects outline a slender torus known as the Arjuna asteroid belt \citep{1993SciN..143..117C,
      1993LPICo.810..266S}. They constitute a peculiar subclass within the near-Earth asteroid (NEA) population as they follow dynamically cold, 
      Earth-like orbits (see for example \citealt{2008MNRAS.386.2031B,2013MNRAS.434L...1D,2015AN....336....5D}). Most Arjunas do not experience 
      resonant behavior with Earth, but the Arjuna orbital domain includes Earth's co-orbital zone that roughly spans the interval of semimajor axis, 
      $a$, (0.994, 1.006)~au with eccentricity, $e$, and values less than $\sim$0.2 (see for example \citealt{2018MNRAS.473.3434D}).

      Although the Arjunas are the closest asteroid population to Earth, the first members of this class were only found a few decades ago 
      \citep{1993Natur.363..704R}. Their small sizes and narrow visibility windows make them difficult to study. Therefore, our current understanding 
      of the origin and evolution of this population is still very limited, despite members of this class sharing the highest economic (mining) and 
      scientific (sample retrieval) interests (see for example \citealt{1997AcAau..41..637S,2013CeMDA.116..367G,2018JSpRo..55...37B}). Moreover, and 
      no less importantly, Arjunas may evolve dynamically into impactors (see for example \citealt{2020AcAau.176..383S}). Slowly but steadily, NEA 
      surveys are uncovering the true extent and complexity of this population.

      Here, we focus on 2023~FY$_{3}$, a recent discovery, the orbit of which might be compatible with present-day co-orbital status, leading to temporary 
      captures and perhaps an impact. This paper is organized as follows. In Sect.~\ref{Data}, we outline the dynamical context of this research and 
      present the data and tools used in our orbital analyses. In Sect.~\ref{Results}, we explore the dynamical evolution of 2023~FY$_{3}$. 
      Details of our observations and their results are given in Sect.~\ref{Observations}. In Sect.~\ref{Discussion}, we discuss our physical and 
      dynamical results. Our conclusions are summarized in Sect.~\ref{Conclusions}.

   \section{Dynamical context, data, and tools\label{Data}}
      Here, we review the theory needed to understand the dynamical results that are presented. Relevant data and tools are also discussed in 
      this section.

      \subsection{Dynamics background}
         Most members of the Arjuna asteroid belt are passing bodies, the mean longitudes of which, relative to Earth, $\lambda_{\rm r}=\lambda-\lambda_{\oplus}$ 
         --- where $\lambda$ and $\lambda_{\oplus}$ are the mean longitudes of the object and Earth, respectively --- circulate or oscillate freely 
         in the interval (0, 360){\degr}. $\lambda=M+\Omega+\omega$, where $M$ is the mean anomaly, $\Omega$ is the longitude of the ascending node, 
         and $\omega$ is the argument of perihelion (see for example \citealt{1999ssd..book.....M}). However, some other members are subjected to the 1:1 
         mean-motion resonance with our planet, going around the Sun in almost exactly one orbital period of Earth, and experiencing temporary and 
         sometimes recurrent libration of $\lambda_{\rm r}$ as this critical angle oscillates about well-defined values (see for example 
         \citealt{1981Icar...48....1D,2002Icar..160....1M,2021MNRAS.507.1640C,2022AJ....163..211Q,2023Icar..39015330D}).

         Among Earth co-orbitals, the most numerous group includes those objects for which the $\lambda_{\rm r}$ librates about 180{\degr} as they roam 
         Earth's Lagrangian point, L$_3$, following horseshoe paths relative to our planet (see for example \citealt{1973ApL....15....1H,
         2012MNRAS.426.3051C,2020MNRAS.496.4420K}). In the classical case, the semi-amplitude of this libration is $<$180{\degr}, but $>$ 90{\degr}. 
         Although the existence of these objects was first considered by \citet{1911MNRAS..71..438B}, it took time to find and confirm 
         the first ones hosted by Earth, 54509 YORP (2000~PH$_{5}$) \citep{2002AGUFM.P11A0352W} and 2002~AA$_{29}$ \citep{2002M&PS...37.1435C}. 
         Additional examples were identified soon after \citep{2004Icar..171..102B}. 

         Minor bodies following horseshoe paths can approach Earth from behind (evening sky) or from the front (morning sky). Sometimes the relative 
         velocity near perigee is close to or under 1~km~s$^{-1}$, leading to temporary captures. Following \citet{2017Icar..285...83F}, if an object 
         manages to complete at least one full revolution around Earth while bound (negative geocentric energy, \citealt{1979RSAI...22..181C}), it 
         becomes a temporarily captured orbiter, and if it does not, a temporarily captured flyby. A similar terminology was used by 
         \citet{1973AJ.....78..316E} within the context of temporary satellite captures by Jupiter and Saturn.

         Temporarily captured small bodies are often referred to as mini-moons. Four such objects have already been identified: 1991~VG 
         \citep{1997CeMDA..69..119T,2018MNRAS.473.2939D}, 2006~RH$_{120}$ \citep{2009A&A...495..967K}, 2020~CD$_{3}$ \citep{2020ApJ...900L..45B,
         2020MNRAS.494.1089D,2020AJ....160..277F,2021ApJ...913L...6N}, and 2022~NX$_{1}$ \citep{2023A&A...670L..10D}. Temporarily captured orbiters 
         --- such as 2006~RH$_{120}$ and 2020~CD$_{3}$ --- have the greatest scientific and commercial value \citep{2018FrASS...5...13J}, and they may 
         be found on a yearly basis by upcoming surveys \citep{2020Icar..33813517F}. In general, Arjuna-type objects are expected to be found in 
         significant numbers by future surveys, such as LSST (see for example \citealt{2023ApJS..266...22S}).

      \subsection{Data, data sources, and tools}
         Asteroid 2023~FY$_{3}$ was found on March 25, 2023, at $G$=18.2~mag by K.~W.~Wierzchos observing for the Catalina Sky Survey 
         (CSS)\footnote{\href{https://www.lpl.arizona.edu/css/css\_facilities.html}{https://www.lpl.arizona.edu/css/css\_facilities.html}} 
         using the 0.68-m Schmidt and a 10K CCD camera \citep{2023MPEC....F..138C}.\footnote{\href{https://www.minorplanetcenter.net/mpec/K23/K23FD8.html}
         {https://www.minorplanetcenter.net/mpec/K23/K23FD8.html}} Table~\ref{elements} shows its latest orbit determination, which is compatible with 
         that of a NEA of the Aten dynamical class. It was retrieved from the Jet Propulsion Laboratory's (JPL) Small-Body Database 
         (SBDB)\footnote{\href{https://ssd.jpl.nasa.gov/tools/sbdb\_lookup.html\#/}{https://ssd.jpl.nasa.gov/tools/sbdb\_lookup.html\#/}} provided by 
         the Solar System Dynamics Group (SSDG, \citealt{2011jsrs.conf...87G,2015IAUGA..2256293G}).\footnote{\href{https://ssd.jpl.nasa.gov/}
         {https://ssd.jpl.nasa.gov/}} The orbit determination is referred to the standard epoch JD 2460000.5 TDB, which is also the origin of time in most 
         calculations presented here. The Earth impact risk computed by JPL's Sentry System\footnote{\href{https://cneos.jpl.nasa.gov/sentry/} 
         {https://cneos.jpl.nasa.gov/sentry/}} \citep{2001DPS....33.4108C,2015DPS....4721409C} for the current orbit determination is low but not zero 
         for impacts in 2114 and 2116.\footnote{\href{https://cneos.jpl.nasa.gov/sentry/details.html\#?des=2023\%20FY3}
         {https://cneos.jpl.nasa.gov/sentry/details.html\#?des=2023\%20FY3}} 
%
%
      \begin{table}
       \centering
       \fontsize{8}{12pt}\selectfont
       \tabcolsep 0.14truecm
       \caption{\label{elements}Values of the heliocentric Keplerian orbital elements and their respective 1$\sigma$ uncertainties for 2023~FY$_{3}$.
               }
       \begin{tabular}{lcc}
        \hline
         Orbital parameter                                 &   & value$\pm$1$\sigma$ uncertainty \\
        \hline
         Semimajor axis, $a$ (au)                          & = &   0.997747$\pm$0.000003         \\
         Eccentricity, $e$                                 & = &   0.0435119$\pm$0.0000007       \\
         Inclination, $i$ (\degr)                          & = &   0.24414$\pm$0.00002           \\
         Longitude of the ascending node, $\Omega$ (\degr) & = &  26.393$\pm$0.003               \\
         Argument of perihelion, $\omega$ (\degr)          & = &  65.680$\pm$0.006               \\
         Mean anomaly, $M$ (\degr)                         & = &  58.718$\pm$0.003               \\
         Perihelion distance, $q$ (au)                     & = &   0.954333$\pm$0.000002         \\
         Aphelion distance, $Q$ (au)                       & = &   1.041161$\pm$0.000003         \\
         Absolute magnitude, $H$ (mag)                     & = &  29.0$\pm$0.4                   \\
        \hline
       \end{tabular}
       \tablefoot{The orbit determination of 2023~FY$_{3}$ is referred to epoch JD 2460000.5 (2023-Feb-25.0) TDB (Barycentric Dynamical Time, J2000.0 
                  ecliptic and equinox), and is based on 177~observations with a data-arc span of 30~d (solution date, May 6, 2023, 05:48:29 PDT). 
                  Source: JPL's SSDG SBDB.
                 }
      \end{table}
%
%

         As of July 9, 2023, and out of 178 known NEAs included in JPL's SBDB with $a\in\left[0.99, 1.01\right]$~au, 2023~FY$_{3}$ has the second 
         lowest value of orbital inclination, 0\fdg24, and the eighth lowest value of $e$, 0.044 (all data referred to epoch JD 2460000.5 TDB). 
         These extreme values place 2023~FY$_{3}$ among the dynamically coldest group of known NEAs close to Earth's path. Among those, it is also the 
         third smallest known, with $H=29$~mag that for a visual albedo of 0.223 (see Sect.~\ref{spec}), corresponding to a size of about 5~m. Minor 
         bodies with such properties are intrinsically difficult to discover. Asteroids that small could be produced through subcatastrophic impacts 
         (see for example \citealt{2007Icar..186..498D}), released from a larger parent asteroid during very close encounters with planets 
         following tidal disruption (see for example \citealt{2014Icar..238..156S}), or due to the action of the 
         Yarkovsky--O'Keefe--Radzievskii--Paddack (YORP) mechanism (see for example \citealt{2006AREPS..34..157B}). However, such properties are also 
         compatible with those of objects with an artificial or lunar-ejecta origin. 
 
         The past and future orbital evolution of an object that has a somewhat uncertain orbit determination and that experiences close approaches to 
         the Earth--Moon system that may lead to an impact has to be studied in statistical terms based on the analysis of results from a 
         representative sample of $N$-body calculations that also take into account the uncertainties in the orbit determination (see for example
         \citealt{2018MNRAS.473.2939D,2020MNRAS.494.1089D}). To this end, we performed $N$-body simulations using a direct $N$-body code written by 
         \citet{2003gnbs.book.....A} that implements the Hermite integration scheme formulated by \citet{1991ApJ...369..200M}. This software is 
         publicly available from the website of the Institute of Astronomy of the University of 
         Cambridge.\footnote{\href{http://www.ast.cam.ac.uk/~sverre/web/pages/nbody.htm}{http://www.ast.cam.ac.uk/$\sim$sverre/web/pages/nbody.htm}} 
         Technical details and relevant results from this code were discussed by \citet{2012MNRAS.427..728D}. Our calculations were carried out in an 
         ecliptic coordinate system with the $X$-axis pointing toward the Vernal Equinox and in the ecliptic plane, the $Z$-axis perpendicular to the 
         ecliptic plane and pointing northward, and the $Y$-axis perpendicular to the previous two and defining a right-handed set. Our physical model 
         included the perturbations by the eight major planets, the Moon, the barycenter of the Pluto-Charon system, and the three largest asteroids, 
         (1) Ceres, (2) Pallas, and (4) Vesta. The input data required to perform the calculations described below were retrieved from JPL's SBDB and 
         SSDG {\tt Horizons} online Solar system data and ephemeris computation
         service,\footnote{\href{https://ssd.jpl.nasa.gov/horizons/}{https://ssd.jpl.nasa.gov/horizons/}} using tools provided by the {\tt Python} 
         package {\tt Astroquery} \citep{2019AJ....157...98G} and its {\tt HorizonsClass}.\footnote{\href{https://astroquery.readthedocs.io/en/latest/jplhorizons/jplhorizons.html}
         {https://astroquery.readthedocs.io/en/latest/jplhorizons/jplhorizons.html}} The retrieved initial positions and velocities (see for example
         Appendix~\ref{Adata}) are based on the DE440/441 planetary ephemeris \citep{2021AJ....161..105P}. 

   \section{Dynamical results\label{Results}}
\begin{figure*}
        \centering
         \includegraphics[width=\linewidth]{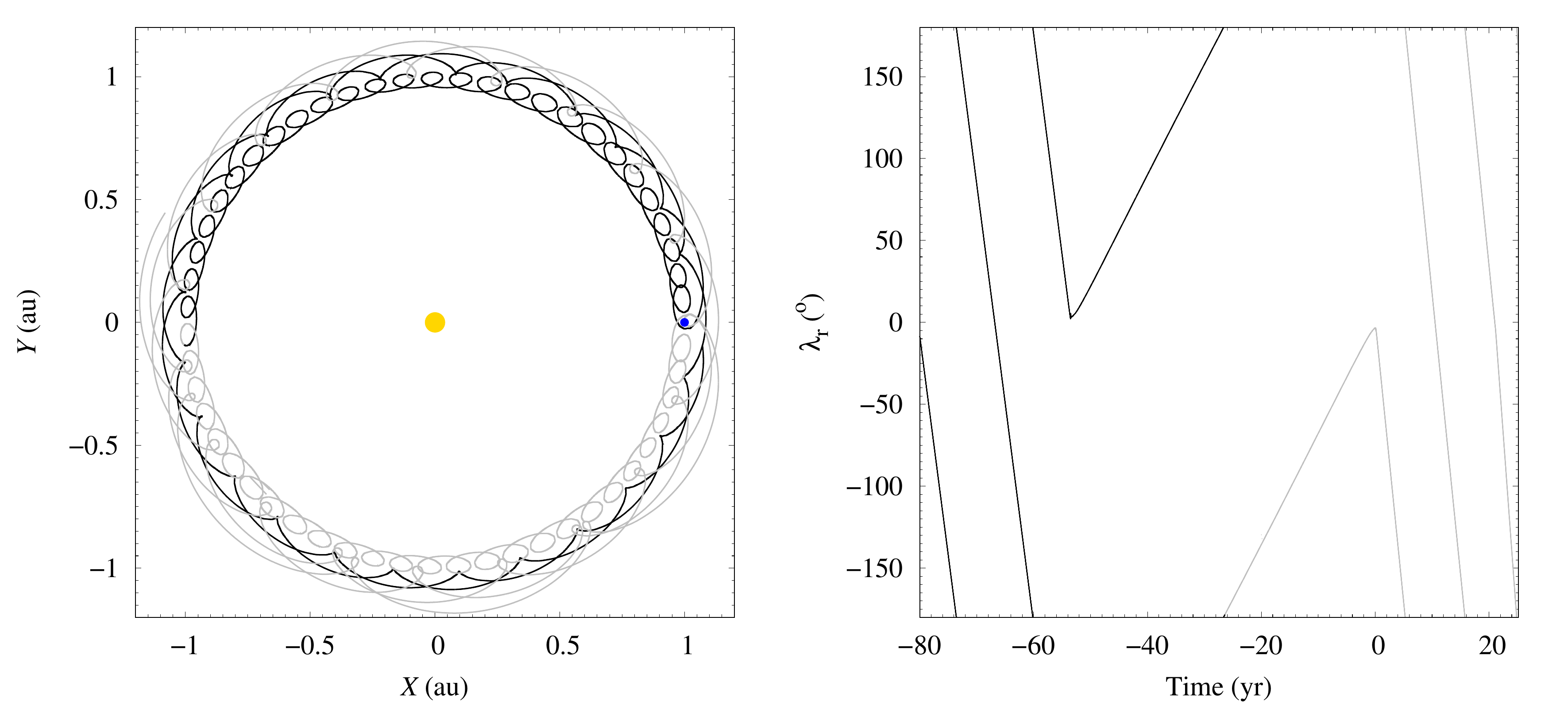}
         \caption{Representative short-term orbital evolution of 2023~FY$_{3}$. \textit{Left panel:} Trajectory in a heliocentric frame of reference 
                  rotating with Earth. \textit{Right panel:} Evolution of the relative mean longitude with respect to Earth, $\lambda_{\rm r}$. The
                  figure corresponds to the evolution of the orbit determination in Table~\ref{elements} in the time interval ($-$80, 25)~yr, with an
                  output cadence of 4.383~h. The curves in black and gray correspond to the same time intervals in both panels.
                 }
         \label{criticalangle}
      \end{figure*}

      The orbit determination of 2023~FY$_{3}$ in Table~\ref{elements} places this object close to the edge of Earth's co-orbital zone and well within 
      the Arjuna orbital realm, $a\in\left[0.99, 1.01\right]$~au (see for example \citealt{2003DDA....34.0615M,2004AGUSM.P11A..04C}). 
      Figure~\ref{criticalangle} shows the short-term evolution of the nominal orbit determination in Table~\ref{elements}. The right panel confirms 
      that 2023~FY$_{3}$ is currently subjected to the 1:1 mean-motion resonance with Earth and exhibits horseshoe-like resonant behavior as the value 
      of $\lambda_{\rm r}$ librates about 180{\degr}. However, the left panel shows that the path followed by 2023~FY$_{3}$ in a heliocentric frame of 
      reference rotating with Earth is not the classical horseshoe described by, for example, \citet{1973ApL....15....1H}, as the object loops around 
      Earth before receding.

%
%
%

\begin{figure}
        \centering
         \includegraphics[width=\linewidth]{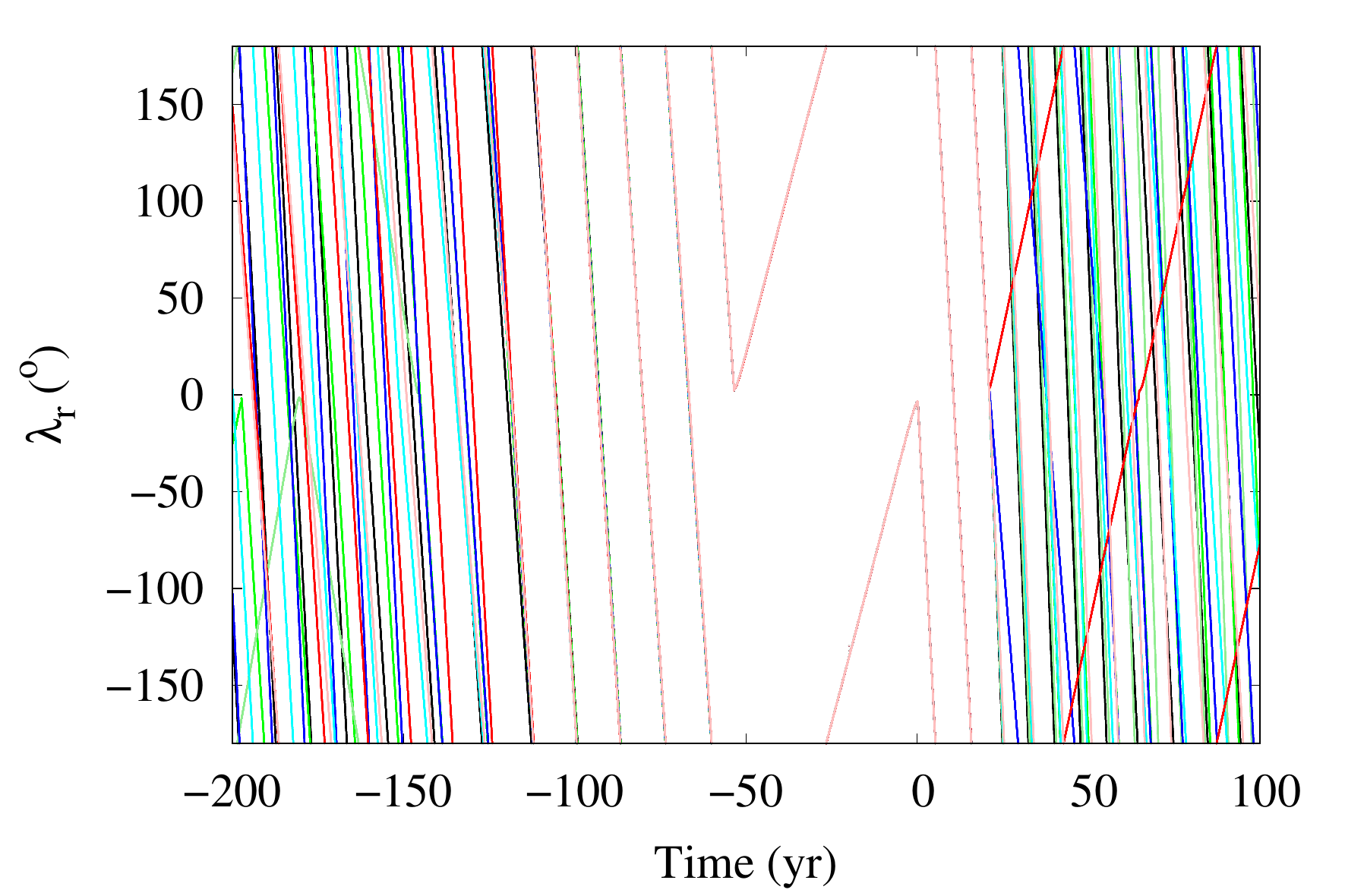}
         \caption{Evolution of the relative mean longitude with respect to Earth, $\lambda_{\rm r}$, of 2023~FY$_{3}$. The figure shows results for 
                  the nominal solution (in black) as described by the orbit determination in Table~\ref{elements} and those of control orbits or 
                  clones with Cartesian state vectors (see Appendix~\ref{Adata}) separated $+$3$\sigma$ (in green), $-$3$\sigma$ (in light green), 
                  $+$6$\sigma$ (in blue), $-$6$\sigma$ (in cyan), $+$9$\sigma$ (in red), and $-$9$\sigma$ (in pink) from the nominal values in 
                  Table~\ref{vector2023FY3}. The time interval ($-$200, 100~yr) is shown. The output cadence is 4.383~h.
                 }
         \label{criticalangleU}
      \end{figure}

      However, considering the uncertainties in Table~\ref{elements}, the robustness of the observed horseshoe-like resonant behavior comes into question. 
      Figure~\ref{criticalangleU} shows the results of the evolution of relevant control orbits that are gradually more separated from the nominal
      one. All the control orbits confirm the current co-orbital status of 2023~FY$_{3}$ as well as its horseshoe-like resonant behavior. The 
      figure also shows that the orbital evolution of 2023~FY$_{3}$ is rather chaotic, as its past becomes unpredictable for times earlier than 1914 
      (over 109~yr ago) but also into the future, beyond 2044 (or about 21~yr from now). This asymmetry is the result of two close encounters with 
      the Earth--Moon system; the future flyby will take place on March 28, 2044, at 0.00925~au (Earth's Hill radius is 0.0098~au) with a relative 
      velocity of 1.29~km~s$^{-1}$.
%
%
%
%

      The uncertainties considered above were assumed to be uncorrelated. While this is a valid assumption when the orbit determination is precisely 
      computed --- in other words, when it is based on a large number of high-quality observations spanning a long time interval --- it is certainly 
      not expected to be true for the orbit in Table~\ref{elements}. In the case of 2023~FY$_{3}$, any two orbital elements may vary with each other 
      and this issue can be properly accounted for by using the covariance matrix that is diagonal when the uncertainties are uncorrelated. In order 
      to quantify the impact of this issue on our results, we carried out additional integrations backward and forward in time of control or clone 
      orbits with initial conditions generated by the Monte Carlo using the Covariance Matrix (MCCM) methodology described in 
      \citet{2015MNRAS.453.1288D}. These synthetic orbits are based on the nominal orbit determination in Table~\ref{elements} with random noise added 
      on each orbital element by making use of the covariance matrix and they are compatible with the observed astrometry. The covariance matrix was 
      retrieved from JPL's SSDG SBDB by using the {\tt Python} package {\tt Astroquery} and its 
      {\tt SBDBClass}\footnote{\href{https://astroquery.readthedocs.io/en/latest/jplsbdb/jplsbdb.html}
      {https://astroquery.readthedocs.io/en/latest/jplsbdb/jplsbdb.html}} class and it is referred to epoch 2460038.5 (April 4, 2023) TDB, which is the 
      start time for the new calculations. The MCCM approach was used to generate initial positions and velocities for 10$^{3}$ control orbits 
      that were evolved dynamically using the direct $N$-body code.
%
%
     \begin{figure}
        \centering
        \includegraphics[width=\linewidth]{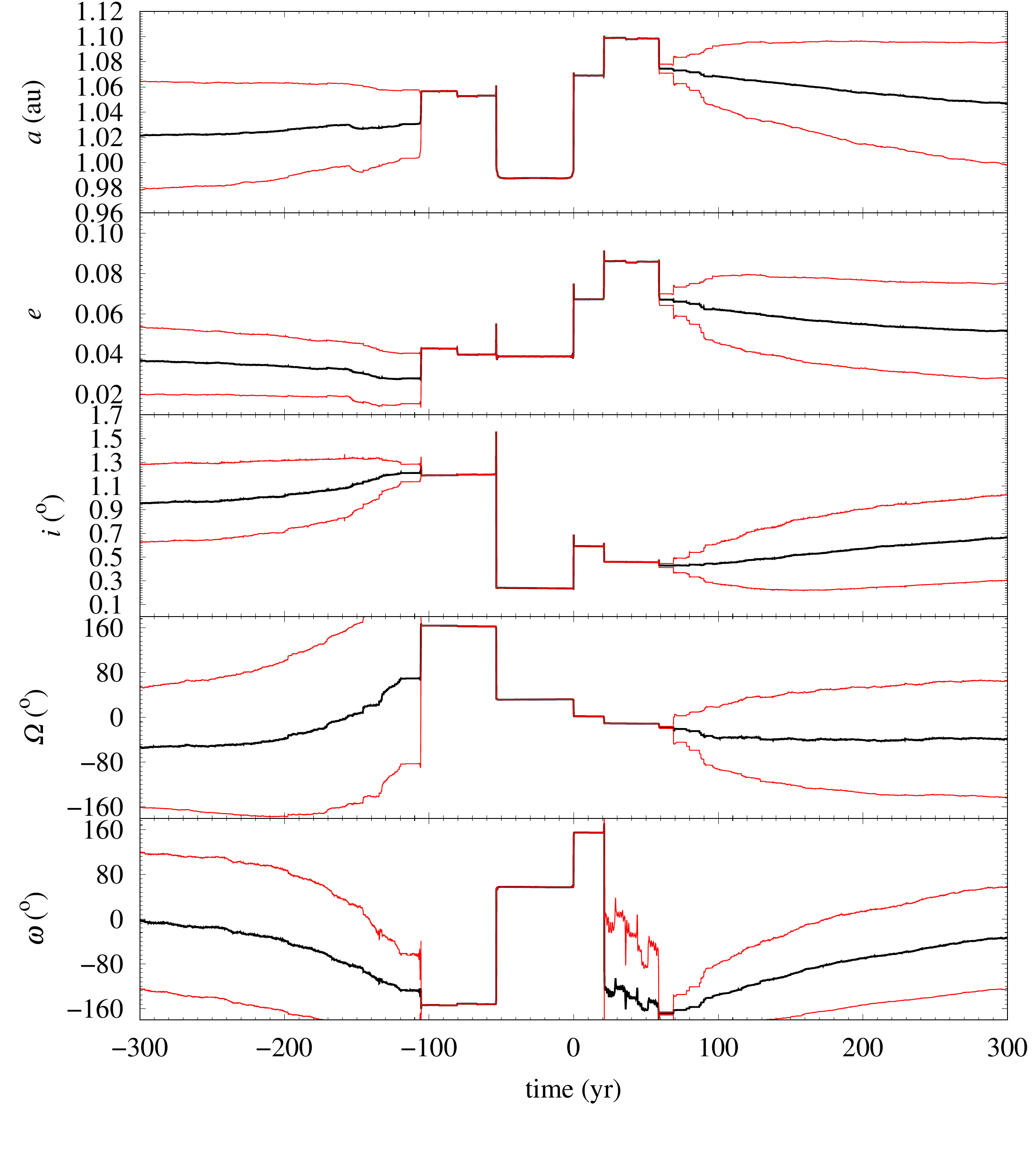}
        \caption{Time evolution of the values of the semimajor axis ($a$, top panel), eccentricity ($e$, second-from-the-top panel), inclination ($i$, third-from-the-bottom panel), ascending node ($\Omega$, second-from-the-bottom panel), and argument of perihelion ($\omega$, bottom panel) of 
                 2023~FY$_{3}$ according to the Monte Carlo covariance matrix (MCCM) approach discussed by \citet{2015MNRAS.453.1288D}. 
                 The panels display the results of the integrations of 10$^{3}$ control orbits with initial positions and velocities generated using the 
                 MCCM method. In black, we display the average evolution of the value of the orbital element and in red we show their respective 
                 ranges, described by the 1$\sigma$ uncertainty or the 16th and 84th percentiles. The output cadence is 0.1~yr. The source of the 
                 input data is JPL's SBDB and Horizons system, and they are referred to epoch 2460038.5 (April 4, 2023) TDB, which is the origin of time 
                 in this figure.
                }
        \label{cova2023FY3}
     \end{figure}
%
%

      Figure~\ref{cova2023FY3} shows the result of the past and future evolution of 2023~FY$_{3}$ when the covariance matrix is factored in. The 
      average evolution is plotted in black and in red we show the range linked to the 1$\sigma$ uncertainty or the 16th and 84th percentiles. The 
      results are consistent with those obtained under the assumption of uncorrelated uncertainties. We confirm that predictions beyond 2044 (CE) are 
      unreliable (see the bottom panel in Fig.~\ref{cova2023FY3}) and that the orbital evolution prior to 1914 (CE) is highly uncertain.

\begin{figure}
        \centering
         \includegraphics[width=\linewidth]{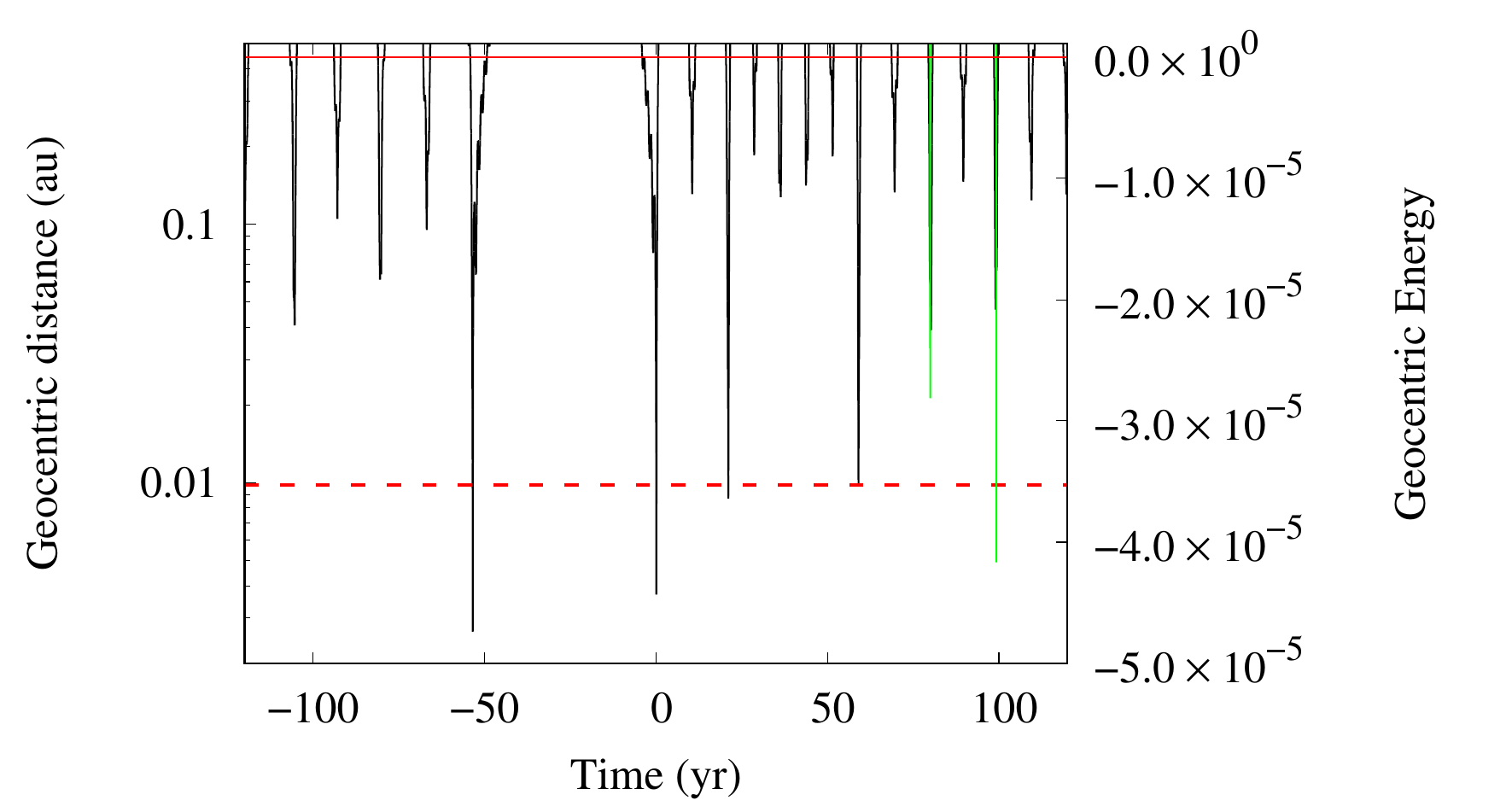}
         \caption{Short-term time evolution of the values of the geocentric distance and energy for the nominal orbit of 2023~FY$_{3}$. The evolution 
                  of the geocentric distance is displayed in black (left-side scale). The value of the Hill radius of Earth, 0.0098~au, is plotted as 
                  a dashed red line. The evolution of the geocentric energy is displayed in green (right-side scale). The zero level is shown as a continuous red line. Captures take place when the value of the geocentric energy becomes negative. The unit of energy is such that the 
                  unit of mass is 1~$M_{\odot}$, the unit of distance is 1~au, and the unit of time is one sidereal year divided by 2$\pi$. The output 
                  cadence is 4.383~h.
                 }
         \label{captures}
      \end{figure}

\begin{figure*}
        \centering
         \includegraphics[width=\linewidth]{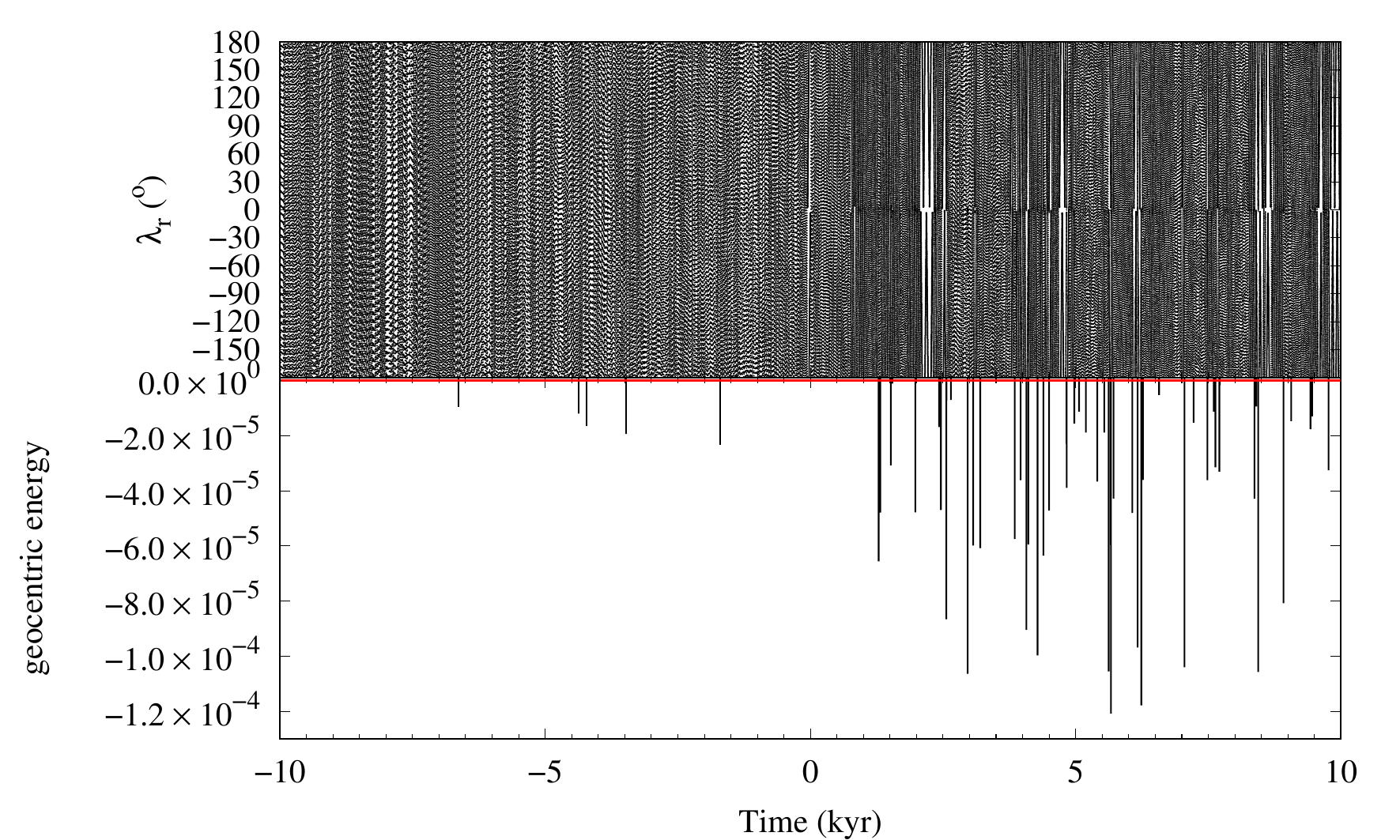}
         \caption{Time evolution of the values of $\lambda_{\rm r}$ and the geocentric energy for the nominal orbit of 2023~FY$_{3}$. \textit{Top 
                  panel:}  $\lambda_{\rm r}$ circulates most of the time although librations are also visible. \textit{Bottom panel:} The evolution of 
                  the geocentric energy shows that recurrent temporary captures are possible. In this figure the output cadence is 0.1~yr and only
                  temporary captures lasting two or more months are visible. Captures take place when the value of the geocentric energy becomes 
                  negative. The unit of energy is such that the unit of mass is 1~$M_{\odot}$, the unit of distance is 1~au, and the unit of time is 
                  one sidereal year divided by 2$\pi$.
                 }
         \label{captures2}
      \end{figure*}

\begin{figure*}
        \centering
         \includegraphics[width=\linewidth]{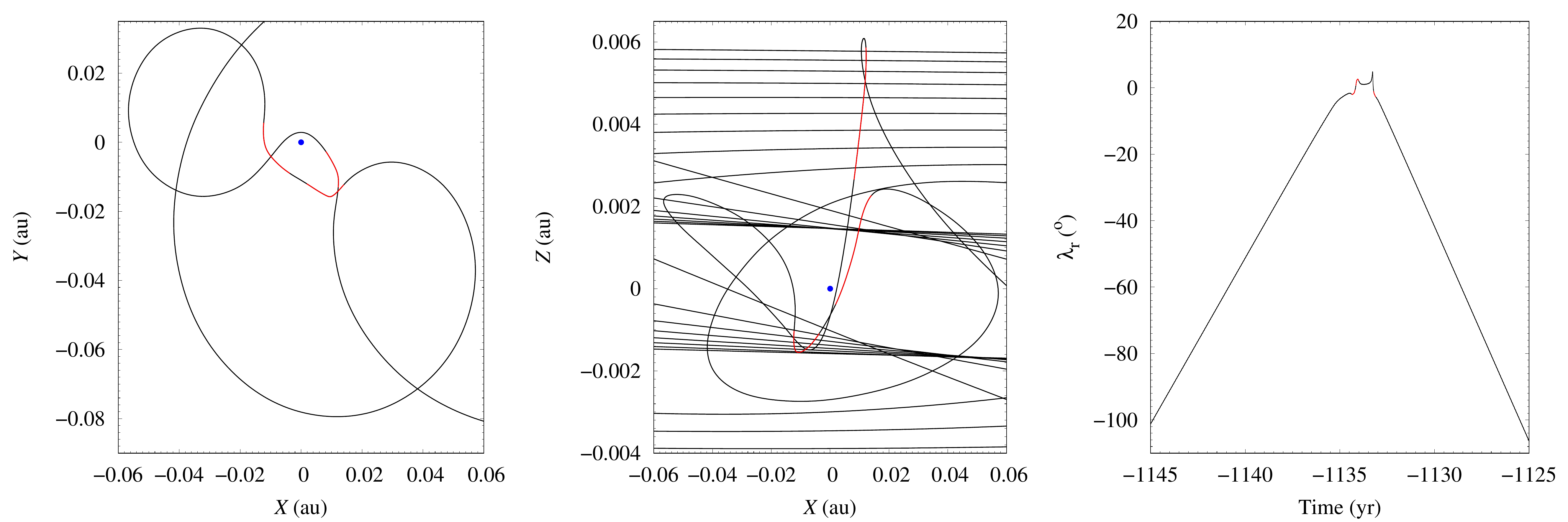}
         \caption{Geocentric trajectory and evolution of $\lambda_{\rm r}$ during one double, temporarily captured flyby. In this example, the curves
                  are plotted in red when the value of the geocentric energy becomes negative. \textit{Left panel:} The geocentric $X-Y$ evolution.
                  \textit{Central panel:} The geocentric $X-Z$ evolution. \textit{Right panel:} $\lambda_{\rm r}$ evolution.     
                 }
         \label{examplecapture}
      \end{figure*}

      We have already pointed out that sometimes the relative velocity of 2023~FY$_{3}$ at perigee near to or inside the Hill radius of Earth can be 
      close to or under 1~km~s$^{-1}$ and this may result in a negative value for the geocentric energy, leading to temporary captures. 
      Figure~\ref{captures} shows that such temporary captures might indeed occur in the near future but the simulated events take place outside the 
      time interval in which predictions based on the current orbit determination are reliable. When longer time intervals are explored, recurrent 
      temporary captures are observed (see Fig.~\ref{captures2}, bottom panel). For most of the simulated time, 2023~FY$_{3}$ remains inside the 
      Arjuna orbital domain but not engaged in 1:1 resonant behavior with Earth (see Fig.~\ref{captures2}, top panel). Although most temporary capture 
      events were of the temporarily captured flyby type like the one shown in Fig.~\ref{examplecapture}, temporarily captured orbiter episodes were 
      also observed.
%
%
%
%
%
%
%
%
%
%
%
%

   \section{Observations\label{Observations}}
      This section describes the different aspects of the observational techniques and data reduction procedures used in our spectroscopic and
      photometric study as well as their results. 

      \subsection{Spectroscopy}
      \label{spec}
          We obtained the visible spectrum of asteroid 2023~FY$_3$ on March 28, 2023 using the Optical System for Imaging and Low Resolution 
          Integrated Spectroscopy (OSIRIS) camera-spectrograph \citep{2000SPIE.4008..623C,2010ASSP...14...15C} at the 10.4~m Gran Telescopio Canarias 
          (GTC), under program GTC31-23A (PI, J.~de~Le{\'o}n). The telescope is located at the El Roque de Los Muchachos Observatory on the island of 
          La Palma (Spain) and is managed by the Instituto de Astrof\'isica de Canarias. The 
          OSIRIS\footnote{\href{http://www.gtc.iac.es/instruments/osiris+/osiris+.php}{http://www.gtc.iac.es/instruments/osiris+/osiris+.php}} 
          instrument was upgraded in January 2023 and is now equipped with a new blue-sensitive, monolithic 4k$\times$4k pixel detector that 
          provides a total field of view (FoV) of 7.8$\times$7.8 arcmin$^2$. We used the 1.2" slit and the R300R grism that gives a resolution of about 
          350 for a 0.6" slit and a dispersion of 7.74~{\AA}~pixel$^{-1}$, covering the wavelength range from 0.48 to 0.92~$\mu$m. The slit was 
          oriented along the parallactic angle and three spectra, with an exposure time of 600~s each, were obtained, with an offset of 10" in the slit 
          direction in between them. The first spectrum was acquired at 21:26~UTC, when the asteroid had an apparent visual magnitude of $m_V=18.4$. 
          The phase angle value and distances to the Sun and to the Earth of the target at the time of the observations were 15.5{\degr}, 1.003~au, and 
          0.0050~au, respectively. We also observed two solar analogue stars from the Landolt catalogue \citep{1992AJ....104..340L}, namely SA98-978 
          and SA102-1081, at a similar airmass to that of the asteroid, in order to obtain its reflectance spectrum. 

\begin{figure}[h!]
        \begin{center}
         \includegraphics[width=1\columnwidth]{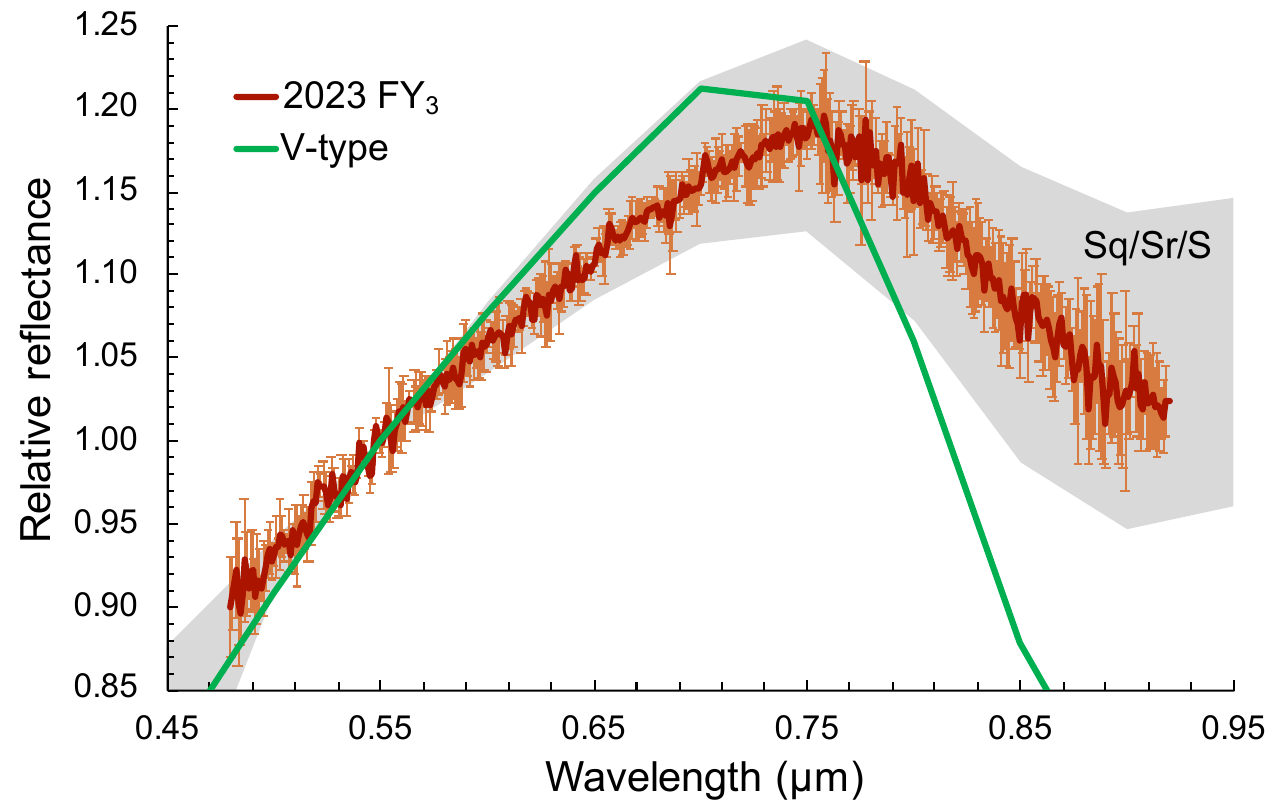}
         \caption{Visible spectrum of asteroid 2023~FY$_3$ obtained with the 10.4~m GTC on the night of March 28, 2023. 
                  Error bars correspond to the standard deviation of the mean. The gray hatched region accounts for the three best taxonomical fits 
                  \citep{2009Icar..202..160D}, in order of increasing $\chi^2$: Sq, Sr, and S-type. We also show the taxon corresponding to V-types 
                  (basaltic asteroids), in green, for comparison purposes.}
         \label{fig:FY3spectrum}
        \end{center}
      \end{figure}

          Data reduction was done using standard procedures. The images were bias and flat-field corrected, and the sky background was subtracted. A 
          one-dimensional spectrum was extracted from the 2D images using an aperture corresponding to the pixel where the intensity decayed to 10\% 
          of the peak value. Wavelength calibration was applied using Xe+Ne+HgAr lamps. The same reduction and extraction procedure was applied to the 
          spectra of the asteroid and the stars. We then divided the asteroid's individual spectra by the spectra of the solar analogues, and the 
          resulting ratios were averaged to obtain the final reflectance spectrum of 2023~FY$_{3}$ shown in Fig.~\ref{fig:FY3spectrum}. The error bars 
          are associated with the standard deviation of the average. 
          
          We used the M4AST\footnote{\href{https://spectre.imcce.fr/m4ast/index.php/index/home}{https://spectre.imcce.fr/m4ast/index.php/index/home}} 
          online tool \citep{2012A&A...544A.130P} to taxonomically classify asteroid 2023~FY$_3$. This tool fits a curve to the spectrum and compares 
          it to the taxons defined by \citet{2009Icar..202..160D} using a $\chi^2$ procedure. As it is shown with a gray hatch in 
          Fig.~\ref{fig:FY3spectrum}, the best three matches in order of increasing $\chi^2$ value correspond to Sq, Sr, and S-type. We also show the 
          V-type taxon (in green) that is typical of basaltic-like composition and that corresponds to the type of material that is most abundant on the 
          surface of the Moon. Therefore, we can confidently say that asteroid 2023~FY$_3$ is an S-type object, which implies that it has a median 
          albedo of $p_V=0.223\pm0.073$, according to \citet{2011ApJ...741...90M}. This albedo value, together with the object's absolute magnitude, 
          provides a diameter of $D\sim$5~m.
%
%
%
%

      \subsection{Time-series photometry\label{lightcurve}}
         Photometric data were obtained on March 28, 2023 with two units of the Two-meter Twin Telescope (TTT), the TTT1 and TTT2 telescopes, located 
         at the Teide Observatory on the island of Tenerife (Canary Islands, Spain). These are two 0.80~m AltAz telescopes with f/4.4 and f/6.8, 
         respectively, which were under commissioning at the time. The observations were made using the QHY411M cameras \citep{2023PASP..135e5001A} 
         installed in one of the Nasmyth ports of both telescopes. They are equipped with 151~Mpixel 3.76~$\mu$m sCMOS sensors, resulting in an 
         effective FoV of 51.4$\times$38.3~arcmin$^2$ and a pixel scale of 0.22"~pixel$^{-1}$ in TTT1 and 33.1$\times$24.7~arcmin$^2$, 
         0.14"~pixel$^{-1}$ in TTT2. An UV/IR-Cut CMOS-optimized filter, nearly equivalent to SDSS g$^\prime+$r$^\prime$, was used. 

         The target had an apparent visual magnitude of $m_V=18.4$ and was moving at a rate of 14"~min$^{-1}$ at the time of observation. The 
         observing run was divided into five 10~min observing blocks with 6.5~s continuous exposures. The median seeing during the observation period 
         was 1.0", so the object appears slightly elongated in the images. Only two of the observing blocks had sufficient signal-to-noise ratios (SNRs) to observe 
         statistically significant variability. Hence, for the time-series photometry, two datasets observed simultaneously with both telescopes 
         between 21:49--21:59 and 22:08--22:18 UT were included.

         Data reduction was done using standard procedures. The images were bias and sky flat-field corrected. Then, the TTT2 images were binned 
         2$\times$2. Aperture photometry was performed using the {\tt Tycho Tracker}\footnote{\href{https://www.tycho-tracker.com/}
         {https://www.tycho-tracker.com/}} software \citep{2020JAVSO..48..262P}. The images were aligned with bicubic interpolation and downsampled by 
         a factor of two for astrometric calibration, performed using the algorithms of {\tt Astrometry.net} \citep{2010AJ....139.1782L}. A fixed circular 
         aperture of 2$\times$FWHM in the first image was used. An outer ring located at 4$\times$FWHM was used to estimate the sky background signal. 
         The same apertures were used for the comparison stars, selected constraining $0.60<(B-V)<0.70$. The initial and final positions of the track 
         were marked manually to prevent uncertainties in the object position coming from the ephemerides.

\begin{figure}[h!]
        \begin{center}
         \includegraphics[width=1\columnwidth]{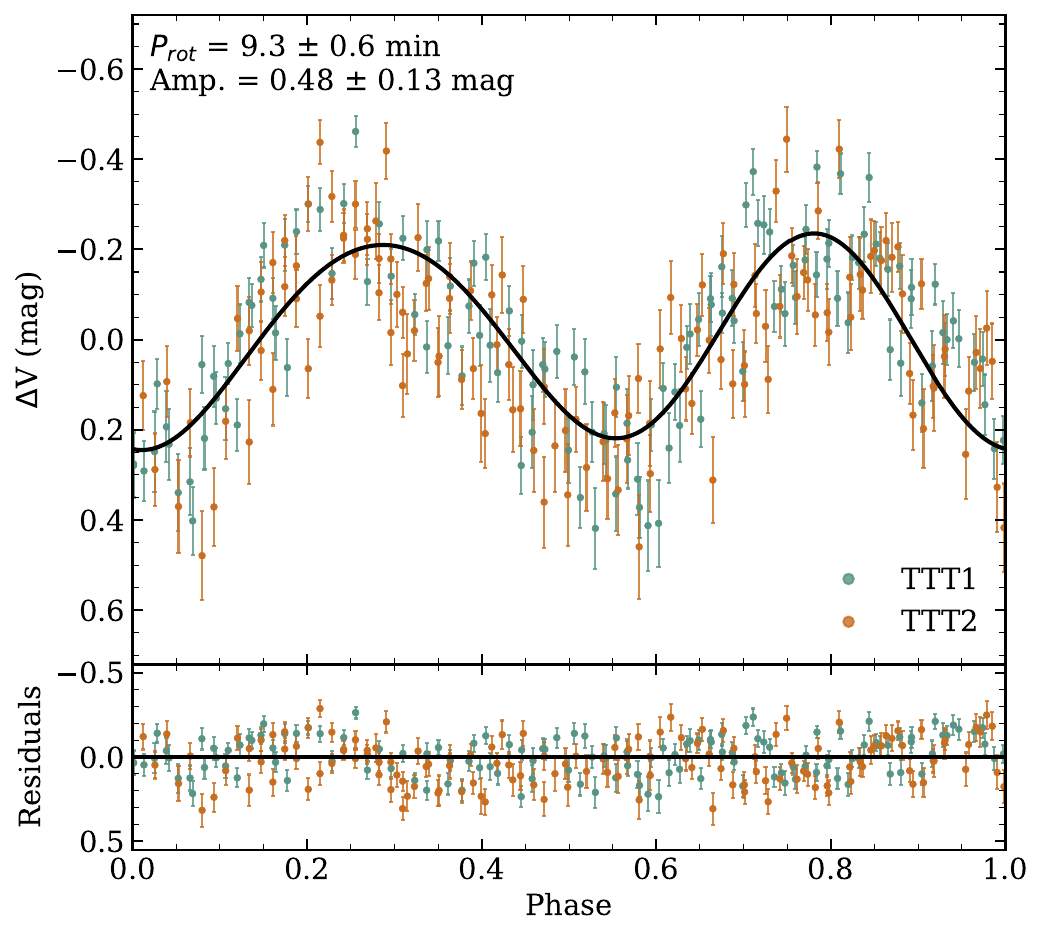}
         \caption{Phased light curve of 2023~FY$_{3}$ derived from photometric measurements obtained by the TTT1 and TTT2 telescopes. The rotation 
                  period and amplitude of the curve are shown at the upper left.} 
         \label{fig:2023FY3_TTT}
        \end{center}
      \end{figure}

         Photometric measurements were extracted and corrected for distance and light-time. The three-term Lomb-Scargle periodogram was obtained, with 
         a wide peak in the power spectrum centered on a period of $P_{\rm rot}=9.3\pm0.6$~min. Aliased peaks at 0.5 and 1.5 times this period are 
         also noticeable. As the uncertainty, 1$\sigma$ of the Gaussian curve fitted to the exponentiated power peak was taken 
         \citep{2018ApJS..236...16V}. The phased light curve is shown in Fig.~\ref{fig:2023FY3_TTT}. The amplitude of the curve computed, considering 
         photometric errors, is 0.48$\pm$0.12~mag. Fast rotation implies intrinsic strength to resist centrifugal disruption 
         \citep{2000Icar..148...12P}. Therefore, 2023~FY$_{3}$ could be a coherent or monolithic body (see for example \citealt{2020MNRAS.495.3990M}). 
         As this asteroid has a nonzero probability of colliding with Earth (see Sect.~2.2), we have to realize that small, denser objects are more 
         likely to survive relatively unaltered the passage through the atmosphere and smash into the ground, producing local destruction. Having 
         knowledge of this information about the bulk strength of an asteroid prior to its impact can help in optimizing mitigation strategies to face 
         eventual damage (see for example \citealt{2023A&A...676A.126P}). However, \citet{2014M&PS...49..788S} suggests that the population of 
         fast-rotating asteroids could include both rubble piles and monolithic boulders, because piles of rubble and dust can be held together by 
         van der Waals forces. Considering this result, 2023~FY$_{3}$ might be made of relatively small grains and dust particles clinging on due to 
         cohesive forces arising from weak electrostatic interparticle attractions that grow in strength for smaller particles as the forces are 
         proportional to the contact surface area.
%
%
%
%

   \section{Discussion\label{Discussion}}
      NEA 2023~FY$_{3}$ is unlikely to have an artificial origin or be lunar ejecta. This conclusion is clear when considering the spectrum in 
      fig.~\ref{fig:FY3spectrum} that corresponds to 2022~NX$_{1}$ (a former mini-moon) and those in fig.~B.1 of \citet{2023A&A...670L..10D} that 
      includes the visible spectra of different artificial objects, and fig.~2 of \citet{2021ComEE...2..231S} that shows the spectrum of 469219 
      Kamo`oalewa (2016~HO$_{3}$) --- a co-orbital that alternates between quasi-satellite and horseshoe resonant states \citep{2016MNRAS.462.3441D} --- that 
      matches those of certain lunar materials. As pointed out above, the question of what 2023~FY$_{3}$'s origin is cannot be addressed by taking 
      into account only its present-day trajectory as described by the orbit determination in Table~\ref{elements} because the reconstruction of its 
      orbital past farther than about 100~yr ago is very uncertain. An alternative approach to answering the question is in the use of an orbital 
      distribution model for NEAs. 

      The NEOMOD Simulator\footnote{\href{https://www.boulder.swri.edu/~davidn/NEOMOD\_Simulator}
      {https://www.boulder.swri.edu/$\sim$davidn/NEOMOD\_Simulator}} \citep{2023AJ....166...55N} was developed by numerically integrating asteroid 
      orbits from sources in the main asteroid belt and calibrating the results on observations of the CSS. Although the model is only strictly 
      applicable to objects with $H\leq28.0$~mag and 2023~FY$_{3}$ has $H=29.0$~mag, it may provide a robust answer to the question if we assume that 
      this object is not a fragment produced through the tidal or rotational breakup of a larger body during a close encounter with the Earth--Moon 
      system. Considering the real number of synthetic NEAs generated by NEOMOD (input option $-1$) in the interval $H\in\left[15.0, 28.0\right]$~mag 
      and the orbits most closely matching the one in Table~\ref{elements}, 2023~FY$_{3}$ may have an origin in the inner region of the main belt with 
      a probability close to 70\% of having been delivered through the $\nu_{6}$ secular resonance and a 20\% chance of coming from the 3:1 
      mean-motion resonance with Jupiter.

      The Arjuna secondary asteroid belt is made of small bodies inserted from the main asteroid belt through resonances or produced in situ via 
      fragmentation processes. These small asteroids follow dynamically cold, Earth-like orbits. Some of them can become co-orbitals, in particular
      horseshoes and quasi-satellites, but also mini-moons. The properties of this population remain poorly understood because relatively large 
      objects are scarce and its members are seldom favorably positioned to be observed from ground-based telescopes. In addition, this region also 
      hosts many objects that are eventually confirmed as active spacecraft or returning space debris. Spectroscopy is critical to identify false
      positives, as pointed out by \citet{2023A&A...670L..10D}, but few Arjunas have spectral information. In addition to the few examples pointed out 
      above, another object with available spectroscopy is the former mini-moon 2020~CD$_{3}$, which has a V-type \citep{2020ApJ...900L..45B}. The 
      relatively short visibility windows of these objects make them challenging spectroscopic targets.  

      Thanks to the study of objects like 2023~FY$_{3}$, the historically blurred picture of the Arjuna secondary asteroid belt is now emerging 
      somewhat clearer and a recent study concludes that Arjunas tend to have the smallest minimum orbit intersection distances (MOIDs) with Earth and 
      the largest values of $H$ \citep{2023A&A...674A..50D}. In this context, 2023~FY$_{3}$ represents a rare window into the subset of the smallest 
      NEAs, but also into the group of ephemeral Earth horseshoes. The spectral type of 2023~FY$_{3}$, when considered within the context of those of 
      other spectroscopically characterized Arjunas (see above), its small size and rapid rotation, and its peculiar dynamics point to a fast-evolving, 
      resonance-driven, diverse Arjuna belt, the study of which may provide insights into the nature of the internal structure of larger, heterogeneous 
      asteroids (their building blocks), and the dynamical context of any planetary defense strategies, as well as the economics of in-space mineral 
      supply procurement.

   \section{Summary and conclusions\label{Conclusions}}
      In this paper, we have presented the results of an observational and numerical study of the Arjuna-type asteroid 2023~FY$_{3}$ --- a small Earth's 
      co-orbital candidate --- that combines several techniques. Spectroscopic observations acquired with the OSIRIS camera spectrograph at the 10.4~m 
      GTC were used to perform a physical characterization of the object; its rotational state was investigated with time-series photometry obtained 
      with QHY411M cameras and two units of the TTT; and its orbital evolution was explored using $N$-body simulations. Our 
      conclusions can be summarized as follows.
      \begin{enumerate}
         \item We show that 2023~FY$_{3}$ is a natural small body with a visible reflectance spectrum consistent with that of an S-type asteroid.
         \item We analyzed its light curve and obtained a rotation period of 9.3$\pm$0.6~min with an amplitude of 0.48$\pm$0.13~mag, typical of small
               monolithic asteroids.
         \item We confirm that 2023~FY$_{3}$ is currently engaged in horseshoe-like resonant behavior with Earth and experiences regular encounters 
               with the Earth--Moon system well inside the Hill radius and at relative velocities close to 1~km~s$^{-1}$.
         \item The orbit determination of 2023~FY$_{3}$ is not robust enough to make reliable predictions of its orbital evolution beyond March 28,
               2044, or reconstruct its dynamical behavior farther than about 100~yr into the past.  
         \item Using the NEOMOD orbital distribution model for NEAs, 2023~FY$_{3}$ may have an origin in the inner region of the main asteroid belt, 
               with a probability close to 70\% of having been delivered through the $\nu_{6}$ secular resonance and a 20\% chance of coming from the 
               3:1 mean-motion resonance with Jupiter. An origin within the Arjuna orbital realm as a result of the fragmentation in situ of a larger
               parent body cannot be discarded.
         \item The analysis of a large sample of control orbits shows that encounters at close range and low relative velocity with the Earth--Moon
               system may have resulted in past temporary satellite captures that could happen again in the future. Although engagements of the 
               temporarily captured flyby type are far more probable, temporary captured orbiter episodes are also possible. 
      \end{enumerate}
      The faint end of the small-body size function is far from being well studied (see for example \citealt{2017AJ....154..170T,2021PSJ.....2...12H}). 
      For 15\% albedo, $H=27.6$~mag is equivalent to a size of 10~m \citep{2021PSJ.....2...12H}. At $H=29.0$~mag, 2023~FY$_{3}$ is one of the smallest 
      known asteroids with both spectroscopic characterization and rotation period determination. Its fast spinning suggests that it could be a 
      coherent body, a single boulder or piece of rubble that came from a larger rubble-pile host, perhaps released by the YORP mechanism. Given its 
      size and dynamical properties, the Arjuna belt may host thousands of objects like 2023~FY$_{3}$. 
      
   \begin{acknowledgements}
      We thank the anonymous referee for a timely, comprehensive, and constructive report. RdlFM and CdlFM thank S.~J. Aarseth for providing one of 
      the codes used in this research and A.~I. G\'omez de Castro for providing access to computing facilities. JdL acknowledges support from the 
      ACIISI, Consejer\'{i}a de Econom\'{i}a, Conocimiento y Empleo del Gobierno de Canarias and the European Regional Development Fund (ERDF) under 
      grant with reference ProID2021010134. JdL also acknowledges financial support from the Spanish Ministry of Science and Innovation (MICINN) 
      through the Spanish State Research Agency, under Severo Ochoa Programme 2020-2023 (CEX2019-000920-S). This work was partially supported by the 
      Spanish `Agencia Estatal de Investigaci\'on (Ministerio de Ciencia e Innovaci\'on)' under grant PID2020-116726RB-I00 /AEI/10.13039/501100011033. 
      Based on observations made with the GTC, installed at the Spanish Observatorio del Roque de los Muchachos of the 
      Instituto de Astrof\'{\i}sica de Canarias, on the island of La Palma. This work is partly based on data obtained with the instrument OSIRIS, 
      built by a Consortium led by the Instituto de Astrof\'{\i}sica de Canarias in collaboration with the Instituto de Astronom\'{\i}a of the 
      Universidad Nacional Aut\'onoma de Mexico. OSIRIS was funded by GRANTECAN and the National Plan of Astronomy and Astrophysics of the Spanish 
      Government. This paper includes observations made during the commissioning of the Two-meter Twin Telescope (TTT) at the IAC's Teide Observatory 
      that Light Bridges, SL, operates on the Island of Tenerife, Canary Islands (Spain). The Observing Time Rights (DTO) used for this research at 
      the TTT have been provided by the Instituto de Astrof\'{\i}sica de Canarias. In preparation of this paper, we made use of the NASA Astrophysics 
      Data System, the ASTRO-PH e-print server, and the MPC data server. 
   \end{acknowledgements}

   \bibliographystyle{aa}

   \begin{appendix}
      \section{Input data\label{Adata}}
         Here, we include the barycentric Cartesian state vector of NEA 2023~FY$_{3}$. This vector and its uncertainties were used to perform some of 
         the calculations discussed above and to generate the figure that displays the time evolution of the critical angle, $\lambda_{\rm r}$ 
         (Fig.~\ref{criticalangleU}). For example, a new value of the $X$ component of the state vector was computed as $X_{\rm c} = X + \sigma_X \ r$, 
         where $r$ is an univariate Gaussian random number, and $X$ and $\sigma_X$ are the mean value and its 1$\sigma$ uncertainty in 
         Table~\ref{vector2023FY3}.
%
%
     \begin{table}
      \centering
      \fontsize{8}{12pt}\selectfont
      \tabcolsep 0.15truecm
      \caption{\label{vector2023FY3}Barycentric Cartesian state vector of 2023~FY$_{3}$: components and associated 1$\sigma$ uncertainties.
              }
      \begin{tabular}{ccc}
       \hline
        Component                         &   &    value$\pm$1$\sigma$ uncertainty                                 \\
       \hline
        $X$ (au)                          & = & $-$8.953431535257687$\times10^{-1}$$\pm$8.55816440$\times10^{-8}$  \\
        $Y$ (au)                          & = &    4.096847623022677$\times10^{-1}$$\pm$5.53879427$\times10^{-7}$  \\
        $Z$ (au)                          & = &    3.456629464841657$\times10^{-3}$$\pm$1.94403735$\times10^{-7}$  \\
        $V_X$ (au/d)                      & = & $-$7.986119960857692$\times10^{-3}$$\pm$8.12817244$\times10^{-9}$  \\
        $V_Y$ (au/d)                      & = & $-$1.568041957560417$\times10^{-2}$$\pm$1.81148808$\times10^{-8}$  \\
        $V_Z$ (au/d)                      & = & $-$4.465281351986577$\times10^{-5}$$\pm$7.02163094$\times10^{-9}$  \\
       \hline
      \end{tabular}
      \tablefoot{Data are referred to epoch JD 2460000.5, which corresponds to 0:00 on February 25, 2023, TDB (J2000.0 ecliptic and equinox). Source: 
                 JPL's {\tt Horizons}.
                }
     \end{table}
%
%

   \end{appendix}

\end{document}